\documentclass[aps, prl, superscriptaddress]{revtex4-2}
\usepackage{graphicx}
\usepackage{booktabs,array}
\usepackage{braket}
\usepackage{subcaption}
\usepackage{caption}
\usepackage{color,soul}
\usepackage{hyperref}
\usepackage{dsfont}

\begin{document}

\author{Rik Broekhoven}
\affiliation{Department of Quantum Nanoscience, Kavli Institute of Nanoscience, Delft University of Technology, 2628 CJ Delft, the Netherlands.}

\author{Curie Lee}
\affiliation{Center for Quantum Nanoscience, Institute for Basic Science (IBS), Seoul 03760, Korea}
\affiliation{Department of Physics, Ewha Womans University, Seoul 03760, Korea}

\author{Soo-hyon Phark}
\affiliation{Center for Quantum Nanoscience, Institute for Basic Science (IBS), Seoul 03760, Korea}
\affiliation{Ewha Womans University, Seoul 03760, Korea}

\author{Sander Otte} 
\affiliation{Department of Quantum Nanoscience, Kavli Institute of Nanoscience, Delft University of Technology, 2628 CJ Delft, the Netherlands.}

\author{Christoph Wolf}
\email{wolf.christoph@qns.science}
\affiliation{Center for Quantum Nanoscience, Institute for Basic Science (IBS), Seoul 03760, Korea}
\affiliation{Ewha Womans University, Seoul 03760, Korea}

\date{\today}
\title{Protocol for certifying entanglement in surface spin systems using a scanning tunneling microscope} 
\maketitle

\section{abstract}
Certifying quantum entanglement is a critical step towards realizing quantum-coherent applications of surface spin systems. In this work, we show that entanglement can be unambiguously shown in a scanning tunneling microscope (STM) with electron spin resonance by exploiting the fact that entangled states undergo a free time evolution with a distinct characteristic time constant that clearly distinguishes it from any other time evolution in the system. By implementing a suitable phase control scheme, the phase of this time evolution can be mapped back onto the population of one entangled spin in a pair, which can then be read out reliably using a weakly coupled sensor spin in the junction of the scanning tunneling microscope. We demonstrate through open quantum system simulations with realistic spin systems, which are currently available with spin coherence times of $T_2\approx$ 300 ns, that a signal directly correlated with the degree of entanglement can be measured at a temperature range of 100$-$400 mK accessible in sub-Kelvin cryogenic STM systems.

\section{introduction}
Recent advances in quantum control of surface spin systems have shown that this platform can be used to design quantum-coherent systems by tailoring the interaction of individual spins using the scanning tunneling microscope (STM) and atom manipulation.\cite{Crommie_Lutz_science_1993b, Yang2017EngineeringSurface, Veldman2021FreeScattering} In such a system, quantum coherent control of single and multiple spins was achieved by electron spin resonance (ESR), which in the STM is facilitated by resonant electric fields.\cite{Baumann2015, Yang2019CoherentSurface, Phark2023Electric-Field-DrivenMagnet,  Phark2023Double-ResonanceSurface, Reina-Galvez2021All-electricEquations, Reina-Galvez2023Many-bodyResonance} When combining the atomic manipulation aspect and quantum coherent control, one can envision that this platform can be used to implement a re-configurable quantum simulator in hardware using only a few atoms and an ESR-STM. The next logical step is to certify entanglement in such as system, which is a strong prerequisite to the study of quantum-coherent phenomena beyond single spin quantum gate operations.\cite{doi:10.1098/rspa.2002.1097, rieffel2011quantum} This, however, is not as straightforward in the ESR-STM since it only allows for time-averaged single spin read-out with long measurement time (ms or kHz),\cite{Bastiaans2018AmplifierFrequencies} compared to the typical time-scale for coherence time ($T_2$) of only several hundred nanoseconds.\cite{Yang2019CoherentSurface, willke2021coherent} Previous works \cite{DelCastillo2023CertifyingESR-STM} have suggested to use the magnetic susceptibility as entanglement witness, however no experimental realization of this idea has yet been shown. Alternatively, one could exploit the fact that entangled states are no longer eigenstates in the Zeeman basis of the constituent spins, and therefore will undergo a time evolution that is distinctively different from the evolution of any non-entangled state. This approach, also called phase reversal tomography,\cite{Wei2020VerifyingCoherences} has been previously shown in phosphorous donor semiconductor qubits.\cite{Stemp2023TomographyQubits} Here, we present a protocol to adapt and optimize this method for ESR-STM, by using the fact that it can be used to probe the free time evolution of spins~\cite{Veldman2021FreeScattering} and has highly sensitive population read out.\cite{Phark2023Double-ResonanceSurface}

\section{Creating and measuring entanglement}
\begin{figure}[h!]
    \includegraphics[width=0.9\textwidth]{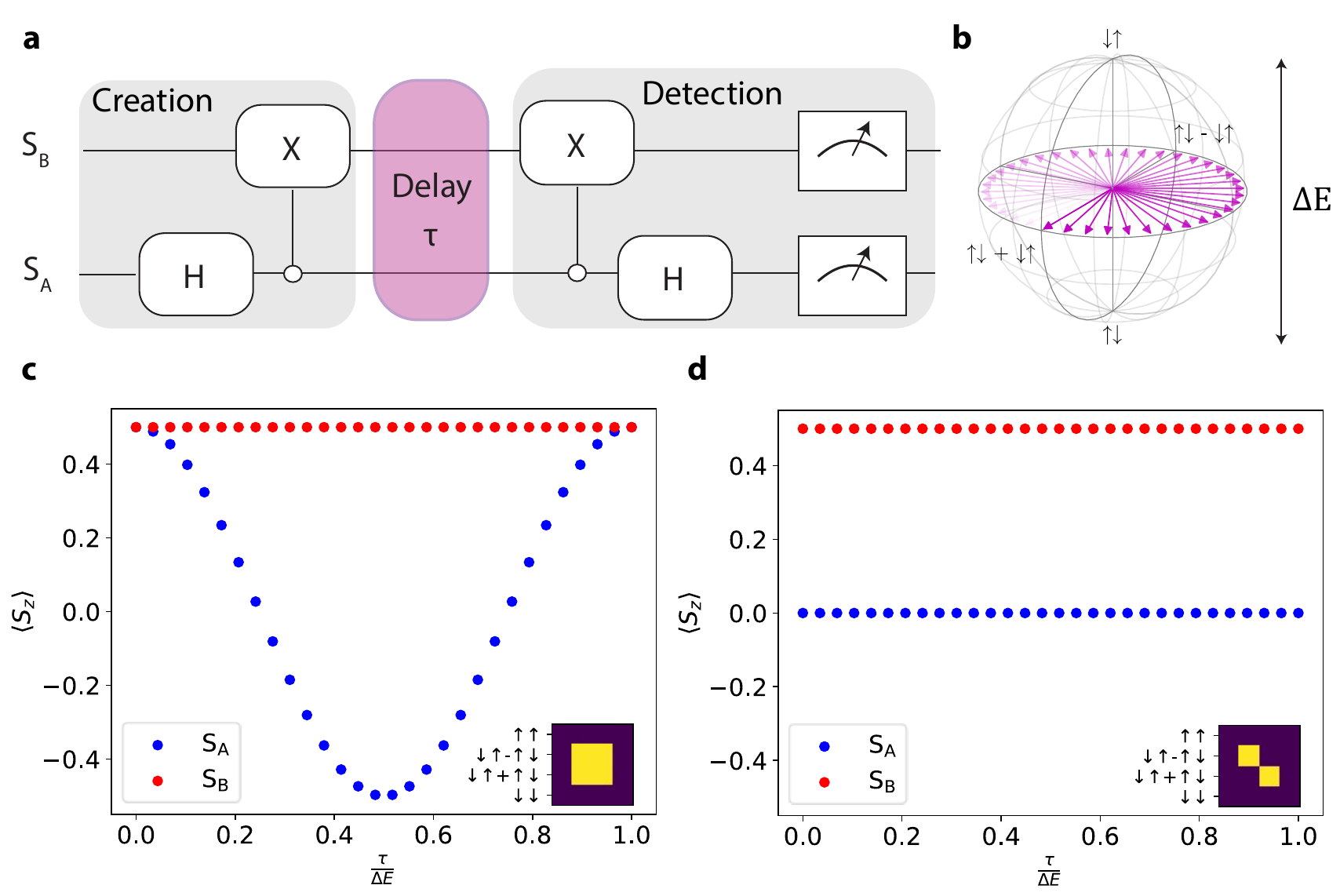}
    \caption{\textbf{Sequence of quantum logic gates to demonstrate entanglement in ESR-STM.} (a) shows the pulse scheme using a quantum gate notation. From left to right this scheme applies a Hadamard gate to spin $A$, a negative CNOT gate, a pulse delay (or phase sweep) gate, followed by an disentangling gate scheme. Finally, both states can be measured to determine their respective populations. (b) the Bloch sphere shows the time-evolution of the entangled state on the equator. (c) expected measurement signal for spin $A$ and spin $B$ when entangled and in contrast according to the density matrix in the inset (d) the same measurement for two spins that are not entangled. }
    \label{entanglementscheme}
\end{figure}

It has been established that ESR-STM provides a universal gate set based on single-spin (or qubit) phase control \cite{Wang2023UniversalSurface} and controlled-NOT gates.\cite{Wang2023AnPlatform} In the following, we will discuss how to create entanglement in a surface spin system and subsequently measure it. We start from two weakly interacting spins, which can be realized in the experiment by using two Ti atoms.\cite{Yang2017EngineeringSurface,Phark2023Double-ResonanceSurface} We require that the interaction between these spins is sufficiently weak so that their combined eigenstates can be written in good approximation as Zeeman product states, e.g. $\ket{\uparrow}_A \otimes\ket{\uparrow}_B =\ket{\uparrow\uparrow}$, where $\uparrow$ ($\downarrow$) denotes the ground (excited state) of each spin and the subscripts $A,B$ label the two spins and $\otimes$ denotes the tensor product. As shown in Fig.~\ref{entanglementscheme}(a), two spins $\ket{\uparrow\uparrow}$ can be entangled by a Hadamard gate $\mathrm{H}$ followed by a negative controlled-NOT gate (CNOT=$\ket{\uparrow}\bra{\uparrow} \otimes \mathds{1} +\ket{\downarrow}\bra{ \downarrow}\otimes \sigma_x$), resulting in an entangled state $\ket{\uparrow\downarrow}+\ket{\downarrow\uparrow}$. In order to detect this entanglement, we can now exploit the fact that the entangled state is not an eigenstate of the Zeeman product basis and thus undergoes a free evolution.\cite{Veldman2021FreeScattering} During this evolution the state picks up a phase at a rate that is proportional to the energy splitting between $\ket{\uparrow\downarrow}$ and $\ket{\downarrow\uparrow}$ (Fig.~\ref{entanglementscheme}(b)). The accumulated phase is distinct from the free evolution of any other non-entangled state and thus allows to uniquely witness the state as being entangled. In particular, maximally correlated states have no accumulated phase.
 We can measure the phase through a Bell state disentanglement measurement, realized by a CNOT followed by a Hadamard, which projects the phase onto one of the two spins followed by read-out of that spin. To be more precise the $\ket{\uparrow\downarrow}+\ket{\downarrow\uparrow}$ is projected upon $\ket{\uparrow\downarrow}$ whereas $\ket{\uparrow\downarrow}-\ket{\downarrow\uparrow}$, which has a phase of $\pi$, is projected upon $\ket{\uparrow\uparrow}$. The full protocol  is illustrated in Fig.~\ref{entanglementscheme}(a). By repeating the scheme with increasing delay times between the entanglement and disentanglement sequences we can probe the full phase accumulation during free evolution. For the maximally entangled state it shows up as a slow (relative to the Larmor frequencies of the individual spins) variation of $\langle S_z\rangle$ as shown in Fig.~\ref{entanglementscheme}(c). This variation can be read-out through the sensor-spin, where $\langle S_z\rangle\propto\Delta I^{ESR}$, i.e. the change in the tunneling current at spin resonance in the ESR-STM experiment.\cite{Phark2023Double-ResonanceSurface} In contrast, the maximally correlated state will result in a flat signal (Fig.~\ref{entanglementscheme}(d)). To directly and unambiguously evidence entanglement, one has to ensure that the measurement gives an oscillation of spin $A$ whereas spin $B$ stays constant.

In a practical implementation, probing free evolution might be slightly disadvantageous when the evolution time is either very short and approaches typical rise and fall times of the signal generator, or very long and rivals the coherence times $T_2$. Fortunately, the effect of free evolution can also be captured by adjusting the phase $\phi$ on the second CNOT gate such that $\phi=\tau/\Delta E$. In the following, we will use such as phase-sweep instead of a delay-time sweep.

\newpage

\section{Implementation}

\begin{figure}[ht]
\includegraphics[width=\textwidth]{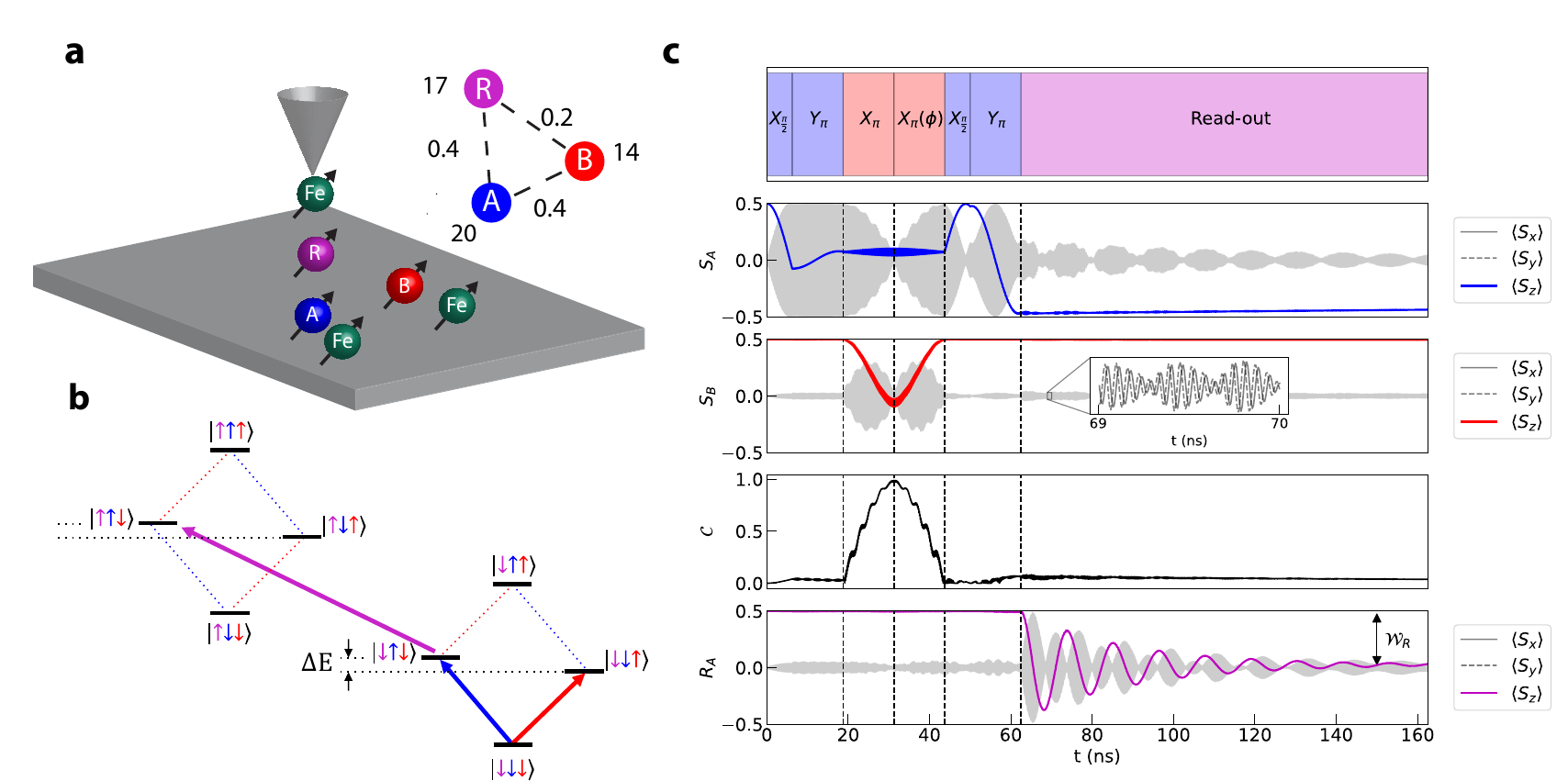}
\caption{\textbf{Two-spin entanglement scheme using sensor spin read-out} (a) Two relatively long-lived spins (A, B) are entangled whilst a third, short-lived sensor spin ($R$) is used for the read-out. Each pair of titanium and iron (Fe) atom serves as a logical qubit in the ESR-STM experiment. (b) energy level diagram showing CNOT (red), Hadamard (blue) spin control and read-out (purple). (c) actual pulse scheme as implemented in the simulations as well as expectation values along $x,y,z$ for each spin involved. The top panel shows the implemented pulse scheme where $X$ and $Y$ represent the rotation axis and the subscript the rotation angle. The next two panels show the time-evolution of spins $A$ and $B$ under driving, followed by the concurrence which serves as direct measure of entanglement in the simulation. The last panel shows the time-evolution of the sensor spin when reading out spin $A$. Idealized parameters were used for clarity: $T$ = 10 mK, $\Omega$ = 0.04 GHz, $T_1^R$ = 20 ns, no relaxation for $A$ and $B$, and Larmor frequencies and exchange couplings are in GHz as indicated in panel (a)} 
\label{sim_setup}
\end{figure}

We will now discuss some details of the implementation. All simulations were carried out using the QuTiP package in the Lindblad formalism using collapse operators parameterized by $T_1$ and $T_\phi$ for energy relaxation and pure dephasing of each spin, respectively (see methods).~\cite{Johansson2012QuTiP:Systems} The total system consists of three spin 1/2 (labeled $A$, $B$, $R$ in Fig.~\ref{sim_setup} (a)), which are exchanged coupled to one another sufficiently weakly so that the state diagram can be written in good approximations as Zeeman product states (details of the system can be found in the methods section). We emphasize that only spins $A$ and $B$ will be the target of this entanglement scheme whilst spin $R$ acts as sensor. The Fe atoms are added in the experiment to provide the local field gradients for driving ESR of the remote spins ($A$ and $B$).\cite{Phark2023Double-ResonanceSurface,Wang2023AnPlatform}

To achieve the desired gate sequence for entanglement, we first combine two rotations (labeled as $X_{\frac{\pi}{2}}, Y_{\pi}$, where $X,Y$ denotes the rotation axis and the subscript the rotation angle) to perform a Hadamard gate and then a single-frequency pulse $X_{\pi}$ to perform a CNOT (Fig.~\ref{sim_setup}(b)). Note that in general in this system a single driving frequency always performs a conditional operation whilst an unconditional NOT gate requires multi-frequency driving.\cite{Wang2023AnPlatform} We found that at low enough temperatures single-frequency driving can be used for all gates due to negligible population in the excited states (Fig.~\ref{sim_setup}(b)). This no longer holds true at elevated temperatures, where excited states can have non-negligible populations. In such a case, the Hadamard gate can result in an admixture of entangled states reflecting the excited state population. To avoid this, we also use single frequency driving for the Hadamard gate, which ensures that only the targeted fraction of population will be entangled, at the loss of overall signal amplitude. We have confirmed that this maximizes the readout of the sensor spin (denoted by $\mathcal{W}_R$ in the following) in our scheme and does not influence the outcome of the entanglement. 

We drive all spins on resonance using a control field of the form $\Omega \cos(\omega_{RF}t+\phi)\hat{\sigma_x}$, with $\Omega$ the Rabi rate, $\omega_{RF}$ the angular radio-frequency resonant with a desired transition, $\phi$ an adjustable phase  and $\hat{\sigma}_x$ the Pauli matrix. It was previously shown that this approach leads to efficient ESR in excellent agreement with the experiment.\cite{Phark2023Double-ResonanceSurface}. For disentanglement we use the same gate sequence but in opposite order whilst matching the initial phase of each subsequent pulse to the phase of the previous pulse. The top 3 plots of Fig.~\ref{sim_setup} (c) show the expectation values for the spin operator $\langle S\rangle$ under these driving fields. We note that the appearance of filled areas is due to crosstalk of the driving frequencies of the pulses and the very fast Larmor precession (10-20 GHz) of each individual spin (see inset), due to the choice that we implemented the simulation in a lab frame of reference. $\phi$ of the second CNOT was chosen such that it mimics half a free evolution in the entangled state resulting in a spin flip of $A$ at the end of the scheme whereas spin $B$ remains unchanged.  At the point where the spin should be entangled the expectation value of $\langle S_z\rangle$ for $A$ and $B$ are 0, indicating that the spins lie at the equator of their Bloch spheres. To further confirm entanglement we also plot the concurrence $\mathcal{C}$, which is bounded by 0 for non-entangled and 1 for maximally entangled states.\cite{Hill1997EntanglementBits} For a bipartite qubit density matrix $\rho_{AB}$, $\mathcal{C}$ is straightforward to calculate and at the point of entanglement the concurrence approaches 1 for the chosen parameter set.  

Read-out of the final target spin states is achieved by a long RF pulse on $R$ conditioned on the spin state to be read out. In this part of the sequence quantum properties like the phase of the pulse play a lesser role, as the coherence of $R$ is known to be limited by the conduction electrons.\cite{Phark2023Double-ResonanceSurface} Fig.~\ref{sim_setup}(b) shows the transition that is driven for read-out of spin $A$. In the STM-ESR experiment a long DC voltage pulse could be used to measure the resulting oscillation as a change in the tunneling current $\Delta I^{\text{ESR}}$. We set a fast decay time ($T_1=20$ ns) for $R$ mimicking this DC pulse and make sure the pulse is relatively long ($100$ ns) such that $R$ quickly reaches the steady state and the signal becomes only dependent on the spin which is read out. Note that we consider this relaxation only during the read-out since for the other parts of the scheme the DC pulse would not be present.
The bottom plot of Fig.~\ref{sim_setup} (c) shows the evolution of the sensor spin during read-out of spin $A$. Since oscillation of $A$ as a function of $\phi$ serves as a witness of entanglement and in Fig.~\ref{sim_setup} (c) $\phi$ is such that this oscillation is at its maximum, we refer to the maximum variation of the sensor spin as the measurement contrast $\mathcal{W}_R$. Due to the nature of the steady state it is at most  half the amplitude of $\langle S_z\rangle$ of $A$. We see that in this case $R$ approaches this value showing that here where the concurrence is 1 the read-out scheme gives a correct output for $\mathcal{W}_R$.  

\section{Results}
\begin{figure}[ht]
\includegraphics[width=\textwidth]{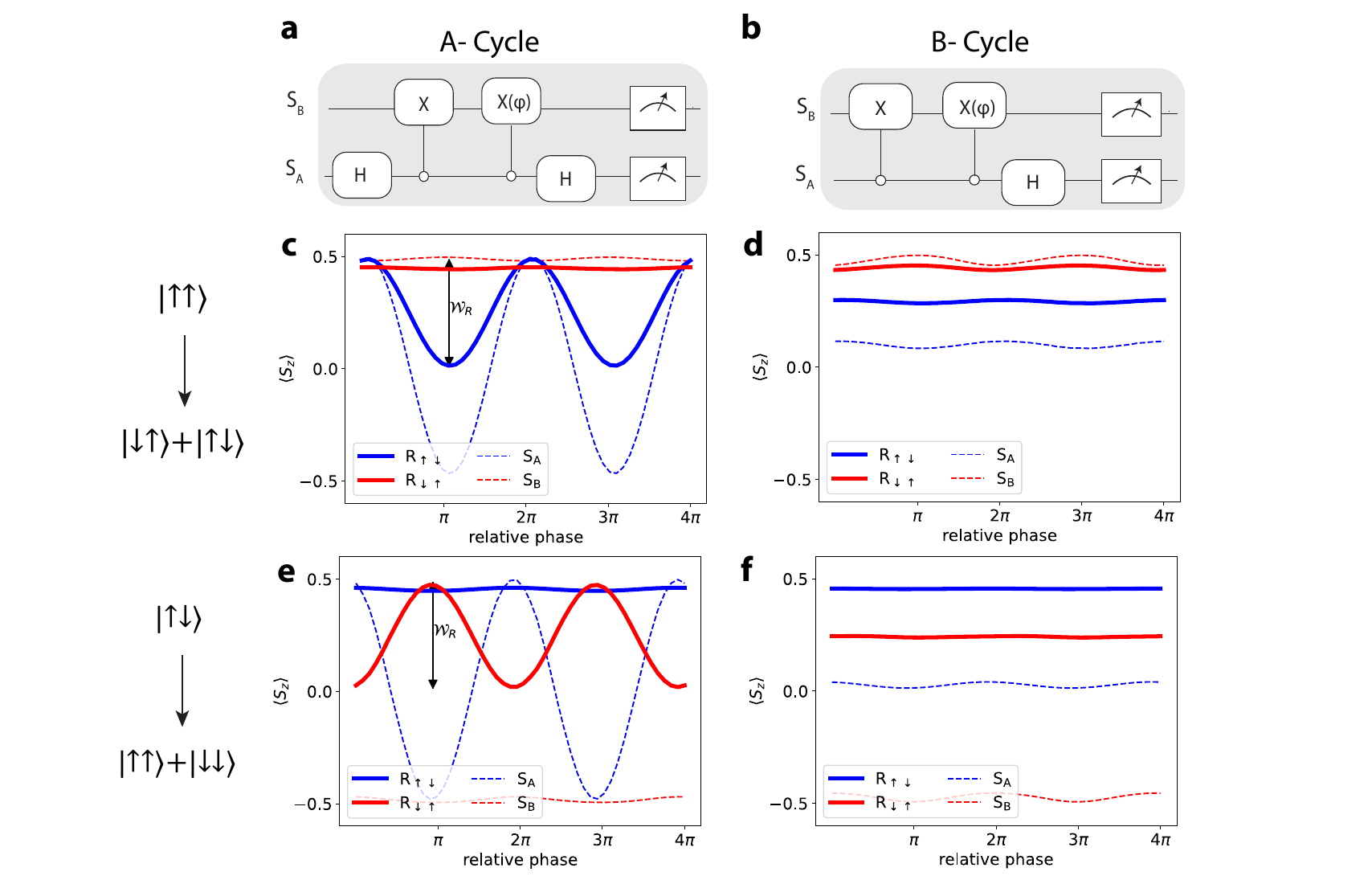}
\caption{\textbf{Simulations of two-qubit entanglement in ESR-STM} showing expected measurement outcomes,  where we compare the expectation values $\braket{S_z}$ on each spin (inaccessible in the experiment) as well as the indirect readout $\mathcal{W}_R$ of these values through the sensor spin. From top to bottom (panels a,c,e) we compare the entangled subspace for different initial states ($\uparrow\uparrow, \uparrow\downarrow$). The panels b, d, f on the right side shows simulations for the same system but for not entangled states (achieved by removing the Hadamard gate), which could serve as empty cycle for the lock-in detection in the ESR-STM experiment. The system parameters are as shown in Fig.~\ref{sim_setup}(a), $T$ = 10 mK, no relaxation for spins A,B, and $T_1$ = 20 ns for the sensor spin during read-out} 
\label{realistiparam_sim}
\end{figure}

To demonstrate the concept, we will first discuss results without any relaxation of spins $A$ and $B$ and at a very low temperature of 10 mK. Larmor frequencies, exchange couplings, and Rabi rates must be chosen such that we stay in the weakly coupled regime while limiting crosstalk, i.e. unwanted driving of other transitions depending on the realistic resonance line widths in the experiment. In addition, we want the Larmor frequencies to be as high as possible to ensure most of the population is in the ground state. Here, we limited the frequency range to $10-20$ GHz which is routinely achieved for single Ti spins on magnesium oxide (MgO) surfaces ($S$ = 1/2) in ESR-STM setups.\cite{Yang2019CoherentSurface, Reale2023ElectricallySurface, Seifert2020Single-atomK} The system parameters are shown in Fig.~\ref{sim_setup}(a), which lists the Larmor frequencies and exchange coupling strengths. In Fig.~\ref{realistiparam_sim}(c-f) we show the results in two ways: first, the variation of each target spin as directly obtained from the density matrix, which serves as evidence of the entanglement but is not accessible with the ESR-STM. Second, we show the expected readout signal $\mathcal{W}_R$, which is a direct observable of the experiment since for the sensor spin $\braket{S_z}\propto \Delta I^{\mathrm{ESR}}$. Contrasting both shows that whilst $\mathcal{W}_R$ is reduced, clearly the signal on the sensor spin directly reflects the spin dynamics of the measurement scheme. We note that in this scheme the phase of the second Hadamard is swept by the equal amount of the free time-evolution, such that $\phi =\omega\tau$, where $\omega$ is the angular frequency associated with the entangled state. Whilst the pulse sequence in Fig.~\ref{realistiparam_sim}(a) can be practically implemented, a real ESR-STM measurement also requires an empty cycle ('B-cycle'), which can be implemented as shown in Fig.~\ref{realistiparam_sim}(b). In this cycle, the background current of the experiment can be measured by simply not entangling the states, which is achieved by removing the Hadamard gate during the entanglement step.
Finally in Fig.~\ref{realistiparam_sim} (e) we show that the method is not limited to the ($\uparrow\downarrow, \downarrow\uparrow$) subspace. Here, we initialise the system in $\downarrow\uparrow$ such that the $H$ and CNOT gate bring the overall target state to $\uparrow\uparrow + \downarrow\downarrow$. Note that here we drive $\downarrow\uparrow$ to $\uparrow\uparrow$ for $H$. The major difference in Fig.~\ref{realistiparam_sim} (e) when compared to Fig.~\ref{realistiparam_sim} (c) is that now the oscillation appears in the read-out of $\ket{\downarrow_A\uparrow_B}$ instead of $\ket{\uparrow_A\downarrow_B}$. This in turn allows to identify all the different Bell states in this system.

\begin{figure}[h]
\includegraphics[width=\textwidth]{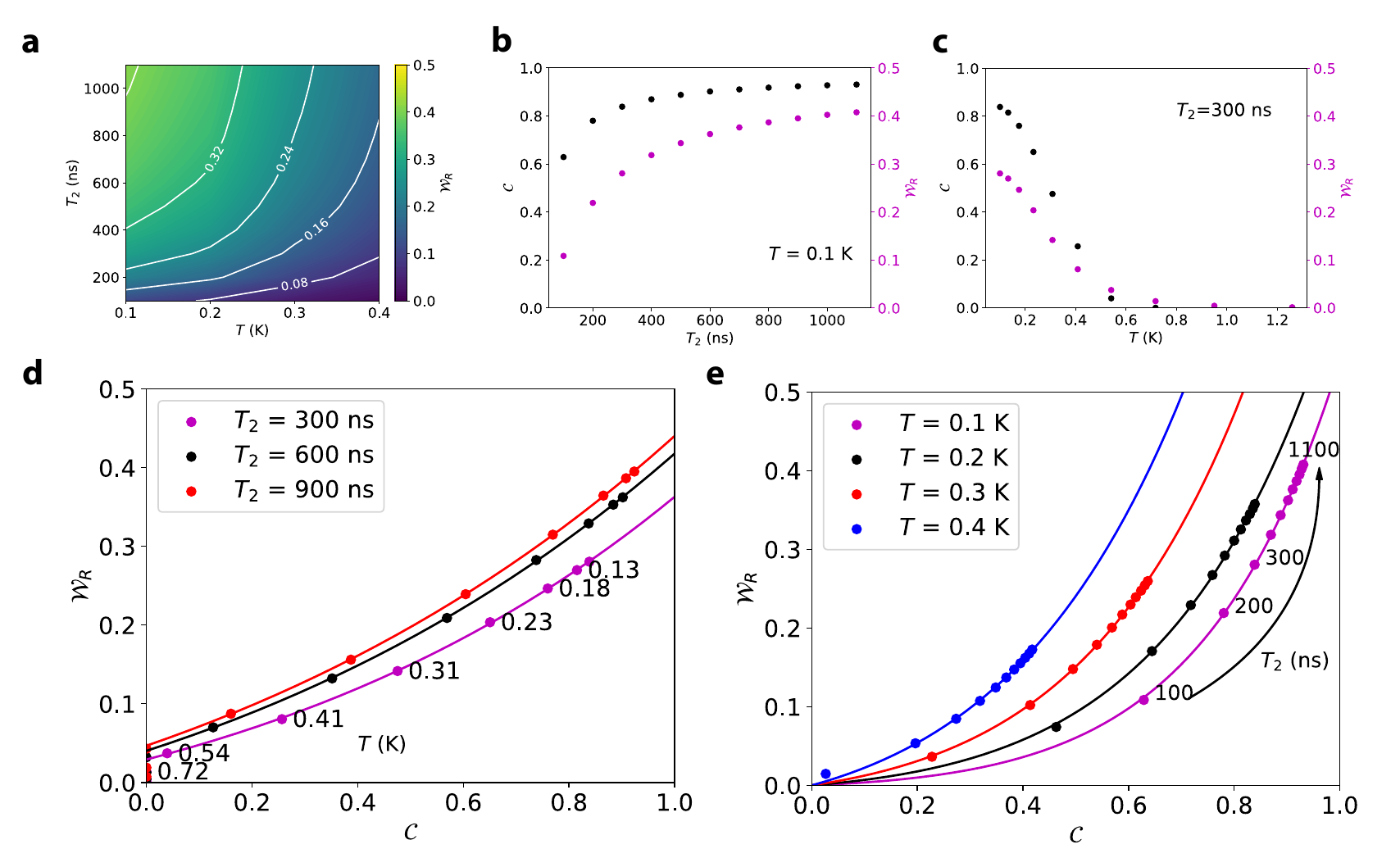}

    \caption{\textbf{Influence of finite lifetime and temperature on the entanglement} (a) shows $\mathcal{W}_R$ as function of temperature and decoherence time $T_2$ (where $T_2=2\cdot T_1$)  (b) slice of (a)  showing $\mathcal{W}_R$ together with the concurrence $C$ for $T=0.1$ K as achievable by dilution refrigerators. (c) slice of (a) showing $\mathcal{W}_R$ together with $C$ for $T_2=300$ ns. Clearly, temperature is a critical factor and the concurrence drops drastically above 0.3 K. (d) relation of $C$ and $\mathcal{W}_R$ for three different $T_2$ with an offset exponential fit (e) relation of concurrence and $\mathcal{W}_R$ for four different temperatures. Solid lines are exponential fits.  The system parameters are as shown in Fig.~\ref{sim_setup}(a)}
        \label{finite_T1_T}
\end{figure}

We now turn to the effect of finite lifetime and elevated temperatures relevant to typical ESR-STM experiments.\cite{Phark2023Electric-Field-DrivenMagnet, Phark2023Double-ResonanceSurface, Wang2023AnPlatform} Previous works have shown that the coherence of Ti spins on two monolayers of MgO deposited on Ag(001) single crystals seem to be lifetime limited such that $T_2=2T_1$, which allows us to discuss the first results without considering additional pure dephasing.\cite{Phark2023Double-ResonanceSurface, Wang2023AnPlatform, Wang2023UniversalSurface}. As can be seen in Fig.~\ref{finite_T1_T}(a)  the $T_2$ time of the two entangled spins (taken here to be identical whilst for the sensor spin $T_1=20$ ns) has a rather modest influence in the experimentally relevant range of $T_2>300$ ns. This is illustrated as well by Fig.~\ref{finite_T1_T} (b), which shows a slice at $T=0.1$ K. Such low temperatures are typically achieved by using a dilution refrigerator equipped ESR-STM which can reach base-temperatures close to 20 mK.\cite{Kim2021SpinField} Clearly, in all cases a $T_2$ of around 300 ns allows for efficient entanglement detection. Temperature is a more critical parameter as becomes apparent in Fig.~\ref{finite_T1_T} (c). Here, we show another slice of Fig.~\ref{finite_T1_T} (a) but now for $T_2=300$ ns. Above 300 mK the concurrence as well as $\mathcal{W}_R$ drastically drop and the concurrences reaches 0 at 700 mK. In the intermediate temperature regime of 400 mK, which can be achieved in a $^3$He-cooled STM system, a small degree of entanglement is achievable, with $\mathcal{C}\approx0.2$ and $\mathcal{W}_R\approx0.1$. The strong temperature dependence is a consequence of reduced population contrast in our system, which is initialized purely by temperature. This means, in turn, that alternative systems where the initialization is achieved by active pumping, might not be as severely limited by temperature. Note that in contrast to $\mathcal{C}$ above 700 mK $\mathcal{W}_R$ is still non-zero meaning that here it is no valid witness of entanglement anymore. We further investigate this in Fig.~\ref{finite_T1_T} (d) where we plot $\mathcal{W}_R$ against $C$ for $T$ sweeps at various $T_2$.  Clearly the observation stands that above 700 mK $\mathcal{W}_R$ is not a valid witness. Fortunately, for temperatures below 700 mK $\mathcal{W}_{R}$ scales with $C$ making a reliable entanglement witness. This relation can best be fitted with a single exponential including an offset, which reflects how at higher temperatures exponentially more population is in unwanted excited states, hereby increasing the effect of crosstalk during the read-out and thus decreasing $\mathcal{W}_R$ compared to its ideal value based on the concurrence. The solid lines in Fig.~\ref{finite_T1_T} (d) are these fits of the form $\mathcal{W}_R=(c+b\mathcal{C})e^{a\mathcal{C}}$ (details of the fitting results can be found in the methods section, Tab.\ref{tab:tab_fig4d}).
Finally, in Fig.~\ref{finite_T1_T} (e)  we plot $\mathcal{W}_R$ against $C$ for $T_2$ sweeps at various temperatures. Again, we see that the dependence can best be fitted exponentially, which here reflects that for lower $T_2$ there is exponentially more decay of the read-out. Solid lines in  Fig.~\ref{finite_T1_T} (e) represent fits of the form $\mathcal{W}_R=b\mathcal{C}e^{a\mathcal{C}}$(details in Tab.\ref{tab:tab_fig4e}).

\begin{figure}[h]
\includegraphics[width=\textwidth]{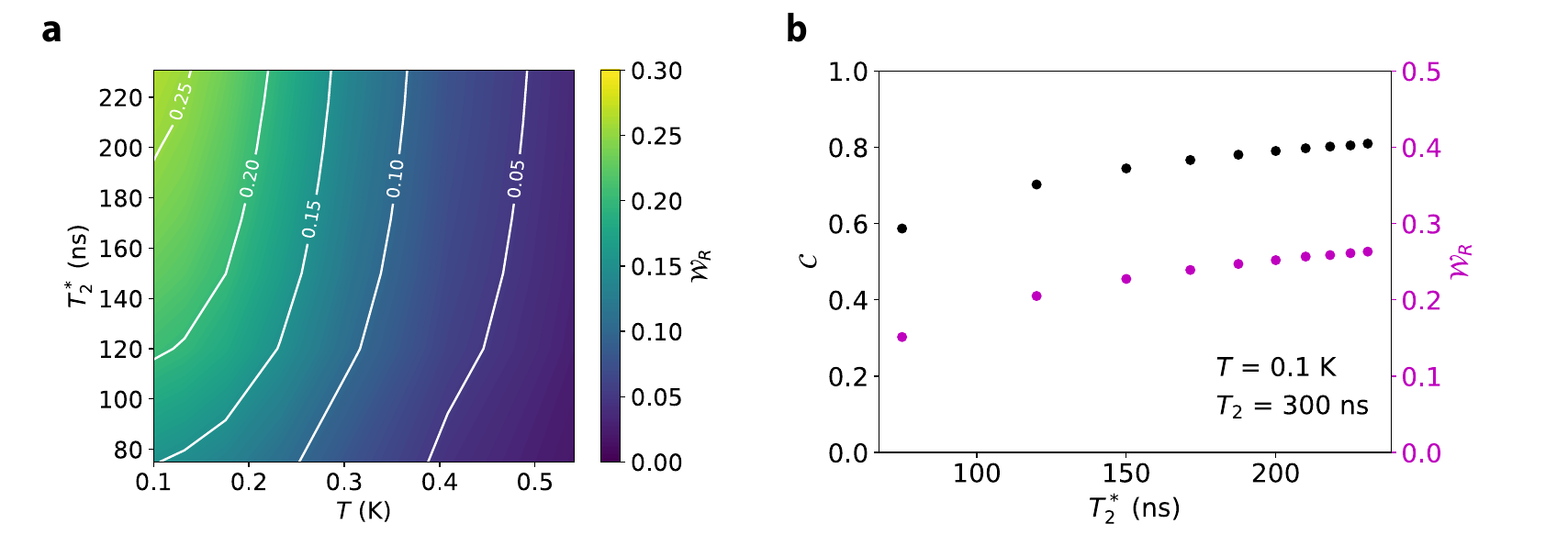}
    \caption{\textbf{Influence of pure dephasing on the entanglement} (a) shows the achievable concurrence and $\mathcal{W}_R$ as function of pure dephasing time $T_\phi$. (b) shows $C$ and $\mathcal{W}_R$ as function of $T_2$ for a fixed $T_2=300$ ns for the two spins and $T=0.1$ K}
    \label{finite_T2}
\end{figure}

Finally, we address systems where coherence is not lifetime limited. In such systems, the coherence of the system is reduced by additional pure dephasing processes, such that an effective coherence time can be defined as $1/T_2^*=1/T_2+1/T_\phi$, with $T_\phi$ being the time constant of the pure dephasing process. In the following, $T_2=2T_1=300$ ns as typical for the experiments \cite{Wang2023AnPlatform}. As show in Fig.~\ref{finite_T2} (a) and (b) even fast dephasing processes with a dephasing time around $T_2^*=75$ ns still allow for sufficient concurrence and $\mathcal{W}_R$. It is not surprising that longer $T_2^*$ times are desirable as this is generally the case in quantum coherent systems, but it is encouraging that in the typical experimental range of $T_2^*\approx 300$ ns \cite{Phark2023Double-ResonanceSurface, Wang2023AnPlatform, Wang2023UniversalSurface} concurrence and $\mathcal{W}_R$ are still relatively high.

\section{conclusion}
In this work we have shown by open quantum systems simulations that two exchange coupled relatively long-lived spins can be entangled and that the entanglement can be directly measured using a third, weakly coupled sensor spin. Our simulations indicate that temperature is critical to achieve high entanglement and $\mathcal{W}_R$, due to the fact that the populations are initialized into thermal equilibrium. Systems that can be initialized more independently from temperature as usually done in optical qubits in trapped ion systems for example, could overcome the strict temperature requirement. For physical spins on surface systems available today, such as the widely studied Ti on MgO/Ag(001), entanglement should be achievable and measurable with $T_2=300$ ns for the quantum spins and $T_1\approx 20$ ns for the sensor spin. High degrees of entanglement $\mathcal{C}> 0.8$ and corresponding read-out can be reached when using a dilution refrigerator at  $T<100$ mK.

\section{Methods}
All calculations were performed by time-evolution for an appropriate amount of time for the entire pulse scheme and the read-out using a converged time step smaller than 8 ps. Following previous works,\cite{Phark2023Double-ResonanceSurface} we modeled each spin as an on-site energy term $2\pi f_{L,i} S_{z,i}$ with $f_{L,i}$ the $i$-th Larmor frequency, and pairwise isotropic exchange coupling terms $J_{i,j} \vec{S}_{i}\vec{S}_{j}$. ESR driving is achieved by applying the necessary single frequency driving terms $\Omega_k\cos(\omega_k (t-t_k^{\textrm{start}})+\phi_k) \sigma_{x,i} (t^{\textrm{start}}_k<t<t^{\textrm{end}}_k)$, with $\omega_k$ the frequency the pulse is send at matching the desired energy transition, $t^{\textrm{start}}_k$ and $t^{\textrm{end}}_k$ the start and end times of the pulse and $\phi_k$ is an adjustable phase. $\Omega_k$ is the on-resonance Rabi rate, $k=1\dots N$ the index of driving frequency terms. The maximum number of driving terms in our simulation was $N\leq 7$. The total system Hamiltonian can be written as follows:\\

\textbf{System Hamiltonian:}

\begin{equation}
H_{tot}=\sum_{i=1}^{3}   2\pi f_{L,i}S_{z,i}+ \sum_{i=1}^{3}  \sum_{j > i}^{3}J_{i,j}\vec{S}_i\vec{S}_j+ \sum_k  \sum_{i=1}^3\Omega_k\cos{(\omega_{ k}(t-t^{\textrm{start}}_k)+\phi_k)\sigma_{x,i}}(t^{\textrm{start}}_k<t<t^{\textrm{end}}_k) 
\end{equation} \\

\textbf{Lindblad equation}: We solved a Lindblad equation for the reduced density matrix $\rho$ of the following form

\begin{equation}
\frac{d\rho}{dt}=-\frac{i}{\hbar}[H_{tot}, \rho]+\sum_l\left( \mathcal{L}_l\rho \mathcal{L}_l^{\dag} -\frac{1}{2}\mathcal{L}_l^{\dag} \mathcal{L}_l\rho -\frac{1}{2}\rho \mathcal{L}_l^{\dag}\mathcal{L}_l\right)
\end{equation} \\

The last term on the right hand side are the collapse operators for our system. We used two sets of collapse operators $\mathcal{L}^{\textrm{Kondo}}+\mathcal{L}^\phi$ to model spin energy relaxation as well as pure dephasing. \\

\textbf{Collapse operators:}
 The first set of collapse operators was defined acting on the coupled 3 spin system in order to model Kondo spin relaxation, known to be the main source of decoherence for these system.\cite{Yang2019CoherentSurface, Ternes2015, Delgado2017}. We arrive at these terms by writing the known rate equation (see for example Eq. 4 of supplementary of  \cite{Loth2010}) in Lindblad form. The operator acting between energy level $m$ and $n$ of the system is

\begin{equation}
\mathcal{L}^{\mathrm{Kondo}}_{m,n} = \sqrt{\sum_l J_l \sum_{s_i, s_f} \bra{m, s_i} \vec{s} \otimes \vec{S_l} \ket{n, s_f}}\frac{\epsilon_{mn}}{1-e^{\epsilon_{mn}/k_B T}} \ket{m}\bra{n}.
\end{equation}

Here the first sum is over the $l$ different atomic spins and the second sum  is over the initial ($s_i$) and final ($s_f$) state of the itinerant electron spin interacting with these spins. $\vec{S}$ and $\vec{s}$ are the respective spin operators. $\epsilon_{mn}$ is the energy difference between $m$ and $n$ of the three spin system. Finally, $J_l$ is the strength of the interaction with each atomic spin. In low temperature approximation it relates to the isolated $l$-th spin relaxation time $T_{1,l}$  and energy of its Larmor frequency $\epsilon_l$ as (see Eq. 69 of \cite{Delgado2017})

\begin{equation}
J_l = \sqrt{\frac{1}{\epsilon_l T_{1,l}}}.
\end{equation}

The second set of operators is for pure dephasing. Here, the standard operators are used relating the pure dephasing rate to the pure dephasing time $T_{\phi,l}$ via the Pauli-z matrix for the $l$-th spin, i.e. $\sigma_{z,l}= \mathds{1} \otimes_1 \mathds{1}\dots \mathds{1} \otimes_l \sigma_z \otimes_{l+1}\mathds{1}\dots$

\begin{equation}
\mathcal{L}^{\phi}_{l} = \sqrt{\frac{1}{2 T_{\phi, l}}} \sigma_{z,l}.    
\end{equation}

\textbf{Read out:}
For read-out long pulses were sent resonant with transitions of $S_R$. The expectation value of $S_R$ was averaged in 16000 time steps for a time of 100 ns. In order to have a converged expectation value a Rabi strength was used double the other strengths used in the scheme.

\textbf{Fitting results:}
The relation between $\mathcal{W}_R$ and $\mathcal{C}$ in Fig.~\ref{finite_T1_T} (d) were best fit using an exponential function of the form $\mathcal{W}_R=(c+b\mathcal{C})e^{a\mathcal{C}}$. The fitting results are reported in Tab.~\ref{tab:tab_fig4d}. The relation between $\mathcal{W}_R$ and $\mathcal{C}$ in Fig.~\ref{finite_T1_T} (e) were best fit using afunction of the form $\mathcal{W}_R=b\mathcal{C}e^{a\mathcal{C}}$. The fitting results are reported in Tab.~\ref{tab:tab_fig4e}.

\begin{table}
\parbox{.45\linewidth}{
\centering
    \begin{tabular}{c||c|c|c}
       $T_2$ (ns) & $a$ & $b$ & $c$\\
       \hline
        300 & 0.68 $\pm$ 0.07 & 0.15 $\pm$ 0.014 & 0.029 $\pm$ 0.027\\
        600 & 0.58 $\pm$ 0.05 & 0.19 $\pm$ 0.011 & 0.040 $\pm$ 0.025 \\
        900 & 0.56 $\pm$ 0.04 & 0.21 $\pm$ 0.011 & 0.047 $\pm$ 0.024\\
    \end{tabular}
    \caption{Fitting results of $\mathcal{W}_R=(c+b\mathcal{C})e^{a\mathcal{C}}$ for the data shown in Fig.~\ref{finite_T1_T} (d). Uncertainties represent the $2\sigma$ confidence interval}
    \label{tab:tab_fig4d}
}
\hfill
\parbox{.45\linewidth}{
\centering
    \begin{tabular}{c||c|c}
       $T$ (K) & $a$ & $b$\\
       \hline
        0.1 & 3.0 $\pm$ 0.06 & (2.7 $\pm$ $0.14
        )\times10^{-2}$ \\
        0.2 & 2.5 $\pm$ 0.07& (5.4 $\pm$ $0.32
        )\times10^{-2}$\\ 
        0.3& 2.2 $\pm$ 0.04 & (9.9 $\pm$ $0.27
        )\times10^{-2}$\\
        0.4 &  1.9 $\pm$ 0.36 & (1.8 $\pm$ $2.5) \times 10^{-1}$\\
    \end{tabular}
    \caption{Fitting results of $\mathcal{W}_R=b\mathcal{C}e^{a\mathcal{C}}$ for the data shown in Fig.~\ref{finite_T1_T} (e). Uncertainties represent the $2\sigma$ confidence interval}
    \label{tab:tab_fig4e}
}
\end{table}

\textbf{Concurrence:}
For concurrence calculation first the partial trace over $S_R$ was taken leaving the reduced matrix in the target spin basis. Then for each entanglement scheme the maximum was reported.   

\textbf{Code availability:}
The underlying code for this study is available and can be accessed via this link 10.5281/zenodo.10528113.

\section*{Author contributions}
RB, SP, and CW conceived the paper, RB and CL performed numerical simulations, all authors contributed to the discussion and the writing of the manuscript.

\section*{Acknowledgments}
The authors thank N. Lorente for his insight related to entanglement in surface spin systems. Further we thank G. Giedke, F. Donati and H. Stemp for discussions. \\
\newline
This work was supported by the Institute for Basic Science (IBS-R027-D1). R. B. and S. O. acknowledge support from the Netherlands Organisation for Scientific Research (NWO Vici Grant
VI.C.182.016).

\section*{Competing interests}
All authors declare no financial or non-financial competing interests. 

\section*{Data availability}
All data generated or analysed during this study are included in 10.5281/zenodo.10530510

\section*{References}
\bibliographystyle{naturemag}
\bibliography{STMentanglementdetection}

\end{document}